\documentstyle[11pt,amssymb]{article}

\let\a=\alpha    
    
  \let\n=\nu

\let\C=\Chi

\def\nn{\nonumber} \def\bd{\begin{document}} \def\ed{\end{document}}
\def\ds{\documentstyle} \let\fr=\frac \let\bl=\bigl \let\br=\bigr
\let\Br=\Bigr \let\Bl=\Bigl
\let\bm=\bibitem
\let\na=\nabla
\let\pa=\partial \let\ov=\overline
\newcommand{\be}{\begin{equation}}
\newcommand{\ee}{\end{equation}}
\def\ba{\begin{array}}
\def\ea{\end{array}}
\def\ft#1#2{{\textstyle{{\scriptstyle #1}\over {\scriptstyle #2}}}}
\def\fft#1#2{{#1 \over #2}}
\def\del{\partial}
\def\vp{\varphi}
\def\st#1{{\scriptstyle #1}}
\def\sst#1{{\scriptscriptstyle #1}}

\def\oneone{\rlap 1\mkern4mu{\rm l}}
\def\td{\tilde}
\def\wtd{\widetilde}
\def\ie{\rm i.e.\ }
\def\dalemb#1#2{{\vbox{\hrule height .#2pt
        \hbox{\vrule width.#2pt height#1pt \kern#1pt
                \vrule width.#2pt}
        \hrule height.#2pt}}}
\def\square{\mathord{\dalemb{6.8}{7}\hbox{\hskip1pt}}}

\def\cramp{\medmuskip = 2mu plus 1mu minus 2mu}
\def\cramper{\medmuskip = 2mu plus 1mu minus 2mu}
\def\crampest{\medmuskip = 1mu plus 1mu minus 1mu}
\def\uncramp{\medmuskip = 4mu plus 2mu minus 4mu}

\newcommand{\ho}[1]{$\, ^{#1}$}
\newcommand{\hoch}[1]{$\, ^{#1}$}
\newcommand{\bea}{\begin{eqnarray}}
\newcommand{\eea}{\end{eqnarray}}
\newcommand{\ra}{\rightarrow}
\newcommand{\lra}{\longrightarrow}
\newcommand{\Lra}{\Leftrightarrow}
\newcommand{\ap}{\alpha^\prime}
\newcommand{\bp}{\tilde \beta^\prime}
\newcommand{\tr}{{\rm tr} }
\newcommand{\Tr}{{\rm Tr} }
\def\0{{\sst{(0)}}}
\def\1{{\sst{(1)}}}
\def\2{{\sst{(2)}}}
\def\3{{\sst{(3)}}}
\def\4{{\sst{(4)}}}
\def\5{{\sst{(5)}}}
\def\6{{\sst{(6)}}}
\def\7{{\sst{(7)}}}
\def\8{{\sst{(8)}}}
\def\n{{\sst{(n)}}}
\def\cA{{{\cal A}}}
\def\cF{{{\cal F}}}
\def\tV{\widetilde V}
\def\tW{\widetilde W}
\def\tH{\widetilde H}
\def\tE{\widetilde E}
\def\tF{\widetilde F}
\def\tA{\widetilde A}
\def\im{{{\rm i}}}
\def\jm{{{\rm j}}}
\def\km{{{\rm k}}}

\def\tY{{{\wtd Y}}}
\def\ep{{\epsilon}}
\def\vep{{\varepsilon}}
\def\R{\rlap{\rm I}\mkern3mu{\rm R}}
\def\bD{{{\bar D}}}
\def\R{{{\Bbb R}}}
\def\C{{{\Bbb C}}}
\def\H{{{\Bbb H}}}
\def\CP{{{\Bbb C}{\Bbb P}}}
\def\RP{{{\Bbb R}{\Bbb P}}}
\def\Z{{{\Bbb Z}}}
\def\bA{{{\Bbb A}}}
\def\bB{{{\Bbb B}}}

\newcommand{\NP}{Nucl. Phys. }
\newcommand{\tamphys}{\it Center for Theoretical Physics,
Texas A\&M University, College Station, TX 77843, USA}
\newcommand{\umich}{\it Michigan Center for Theoretical Physics,
University of Michigan\\ Ann Arbor, MI 48109, USA}
\newcommand{\upenn}{\it Department of Physics and Astronomy,
University of Pennsylvania\\ Philadelphia,  PA 19104, USA}
\newcommand{\SISSA}{\it  SISSA-ISAS and INFN, Sezione di Trieste\\
Via Beirut 2-4, I-34013, Trieste, Italy}

\newcommand{\ihp}{\it Institut Henri Poincar\'e\\
  11 rue Pierre et Marie Curie, F 75231 Paris Cedex 05}

\newcommand{\damtp}{\it DAMTP, Centre for Mathematical Sciences,
 Cambridge University\\ Wilberforce Road, Cambridge CB3 OWA, UK}
\newcommand{\itp}{\it Institute for Theoretical Physics, University of
California\\ Santa Barbara, CA 93106, USA}

\newcommand{\auth}{M. Cveti\v{c}\hoch{\dagger}, G.W. Gibbons\hoch{\sharp},
H. L\"u\hoch{\star} and C.N. Pope\hoch{\ddagger}}

\begin{document}
\begin{flushright}
\hfill{DAMTP-2001-25}\ \ \ {CTP TAMU-18/01}\ \ \ {UPR-938-T}\ \ \
{MCTP-01-25}\ \ \ {NSF-ITP-01-48}\\
{May 2001}\ \ \
{hep-th/0106026}
\end{flushright}


\begin{center}
{ \large {\bf  Supersymmetric M3-branes and $G_2$ Manifolds}}

\vspace{5pt}
\auth

\vspace{3pt}
{\hoch{\dagger}\upenn}

\vspace{3pt}
{\hoch{\sharp}\damtp}


\vspace{3pt}
{\hoch{\star}\umich}

\vspace{3pt}
{\hoch{\ddagger}\tamphys}

\vspace{3pt}
{\hoch{\dagger}\itp}

\vspace{3pt}

\underline{ABSTRACT}
\end{center}

We obtain a generalisation of the original complete Ricci-flat metric
of $G_2$ holonomy on $\R^4 \times S^3$ to a family with a non-trivial
parameter $\lambda$.  For generic $\lambda$ the solution is singular,
but it is regular when $\lambda=\{-1,0,+1\}$.  The case $\lambda=0$
corresponds to the original $G_2$ metric, and $\lambda =\{-1,1\}$ are
related to this by an $S_3$ automorphism of the $SU(2)^3$ isometry
group that acts on the $S^3\times S^3$ principal orbits.  We then construct
explicit supersymmetric M3-brane solutions in $D=11$ supergravity, where the
transverse space is a deformation of this class of $G_2$ metrics.
These are solutions of a system of first-order differential equations
coming from a superpotential.  We also find M3-branes in the deformed
backgrounds of new $G_2$-holonomy metrics that include one found by
A.~Brandhuber, J.~Gomis, S.~Gubser and S.~Gukov, and show that they 
also are supersymmetric.


\pagebreak
\setcounter{page}{1}

\vfill\eject

\section{Introduction}

     Recently the study of M-theory on spacs of $G_2$ holonomy has
attracted considerable attention. In particular, it has been proposed
that M-theory compactified on a certain singular seven-dimensional
space with $G_2$ holonomy might be related to an ${\cal N}=1$, $D=4$
gauge theory \cite{acharya,amv,wit-talk,av,aw} that has no conformal
symmetry.  (See also the recent papers
\cite{gomis,en,kacmcg,gutpap,kkp}.)  The quantum aspects of M-theory
dynamics on spaces of $G_2$ holonomy can provide insights into
non-perturbative aspects of four-dimensional ${\cal N}=1$ field
theories, such as the preservation of global symmetries and phase
transitions.  For example, Ref. \cite{aw} provides an elegant
exposition and study of these phenomena for the three manifolds of
$G_2$ holonomy that were obtained in \cite{brysal,gibpagpop}.

    Studying the classical geometry of eleven-dimensional supergravity
on spaces with $G_2$ holonomy provides a starting point for
investigations of $D=4$ vacua in M-theory.  Recently, in
\cite{clpmassless}, new configurations in M-theory were found, which
describe M3-branes with a (3+1)-dimensional Poincar\' e invariance on
the world-volume.  These configurations arise as solutions of $D=11$
supergravity in which the 4-form field is non-vanishing in the
seven-dimensional transverse space, which is deformation of the a
Ricci-flat metric of $G_2$ holonomy.

   The transverse spaces for M3-branes obtained in \cite{clpmassless}
were deformations of the three explicitly known metrics of $G_2$
holonomy that were found in \cite{brysal,gibpagpop}.  The original
metrics are all of cohomogeneity one: two have principal orbits that
are $\CP^3$ or $SU(3)/(U(1)\times U(1))$, viewed as an $S^2$ bundle
over $S^4$ or $\CP^2$ respectively, and the third has principal orbits
that are topologically $S^3\times S^3$.  The M3-branes on these
deformed $G_2$ holonomy spaces arise from 
first-order differential equations derivable from a
superpotential, which in fact are the integrability conditions for 
the existence of a Killing spinor.  Thus the M3-branes obtained 
in \cite{clpmassless} are supersymmetric.  

    A number of issues related to the properties of $G_2$ holonomy
spaces invite further study.  For example, it is of interest to find
generalisations of the original three complete metrics of $G_2$
holonomy that were obtained in \cite{brysal,gibpagpop}.  Of particular
interest is to find families of $G_2$ metrics on these manifolds that
have non-trivial parameters (\ie not merely the scale parameter that
any Ricci-flat metric has).  Such families of metric, when reduced on
a circle could, for example, provide a connection between the
six-dimensional metrics on the deformed and resolved six-dimensional
conifolds \cite{dloc}, thus providing insights into the relation
\cite{amv} between M-theory on these classes of $G_2$ holonomy
metrics, and Type IIA string theory on the corresponding
six-dimensional special holonomy spaces.\footnote{For 8-dimensional
spaces of $Spin(7)$ holonomy an explicit family of non-singular 
generalisations has
been found recently \cite{cglpspin7,cglpspin7m}, which has such a
non-trivial parameter. They provide a connection between ${\cal N}=1$
$D=3$ vacua of M-theory, and Type IIA string theory on $G_2$ holonomy
manifolds with $\R^3\times S^4$ topology.}  Recently, in
Ref. \cite{bggg}, a new class of $G_2$ holonomy metrics with such
properties was constructed.

    In section 2 we construct a new two-parameter family of metrics of
$G_2$ holonomy, on the manifold of $\R^4 \times S^3$ topology.  We do
this by starting from a rather general ansatz for metrics of
cohomogeneity one on manifolds whose principal orbits are $S^3$
bundles over $S^3$, which is a generalisation of the ansatz for the
original $G_2$ metrics of this topology in \cite{brysal,gibpagpop}.
The general ansatz involves nine functions of the radial coordinate,
which parameterise homogeneous squashings of the base and fibre
3-spheres.  We have not found a superpotential formulation for the
Einstein equations in this most general case, but by making a
specialisation to a 3-function ansatz that is spherically symmetric in
the base and the fibres, we have found a superpotential and hence
first-order equations.  We show that these equations are in fact the
integrability conditions for the existence of a single
covariantly-constant spinor, and hence for $G_2$ holonomy.  The
first-order equations can be solved explicitly, yielding Ricci-flat
metrics with two parameters, one of which is the (trivial) scale size,
while the other, which we call $\lambda$, is non-trivial and
characterises genuinely inequivalent metrics.\footnote{There is also a
third completely trivial parameter corresponding to a constant shift
of the radial coordinate.}  For a generic value of the parameter
$\lambda$ these metrics are singular, but they are regular for three
discrete values, namely $\lambda=\{-1,0,+1\}$.  The case $\lambda=0$
corresponds to the original metric of $G_2$ in
\cite{brysal,gibpagpop}, while the cases $\lambda =\{-1,1\}$ are
related to $\lambda =0$ by an $S_3$ (permutation) automorphism of the
$SU(2)^3$ isometry group.\footnote{The discrete $S_3$ coordinate
transformations (for details see section 2.3) were found in \cite{aw}
as a triality symmetry among the three regular $G_2$ holonomy metrics
with important consequences for the phase diagrams of the field theory
associated with these three branches.}  The $S_3$ 
action restricts the parameter space of these metrics to a
``fundamental domain'' $0\le \lambda\le \ft13$.  However, when viewed
as a reduction on a specific circle, M-theory on the three regular
solutions is related to the type IIA theory on a deformed conifold and
two types of resolved conifold respectively.

    In section 3 we consider a different specialisation of the
nine-function metric ansatz introduced in section 2, in which six metric
functions remain.  For this particular truncation a superpotential can
be found, and we show that the resulting first-order equations are the
integrability conditions for a covariantly-constant spinor, and hence
$G_2$ holonomy.  A specialisation in which two of the three directions
on the base and fibre 3-spheres are treated on an equivalent footing
yields the ansatz and first-order system introduced in \cite{bggg}.
We discuss how this is related by dimensional reduction to the metric
on the six-dimensional Stenzel manifold.  A further specialisation of
the six-function ansatz to a spherically-symmetric one with just two
remaining functions yields a first-order system whose solution is the
original metric of $G_2$ holonomy on the $\R^4$ bundle over $S^3$,
expressed now in a manifestly $\Z_2$-symmetric fashion.

    Another aim of the paper is to construct further examples of
supersymmetric M3-branes.  In section 4 we extend the
methods introduced in \cite{clpmassless} to obtain the equations of
motion for M3-branes in the background of deformations of the new
$G_2$ metrics that we constructed in section 2.  We show that these
equations admit a superpotential formulation, and we obtain the
general M3-brane solution for these configurations.  We also obtain
the analogous system of equations for M3-branes in the background of
deformations of the new $G_2$ ansatz introduced in \cite{bggg}, which
we discussed in section 3.  Again, we find a superpotential and
associated first-order equations.  In section 5 we show that 
these first-order equations are the integrability conditions for the
existence of a Killing spinor.  Thus the new M3-brane
solutions are supersymmetric.  

\section{A class of metrics of $G_2$ holonomy}

\subsection{The ansatz and Einstein equations}

   An approach to looking for metrics of $G_2$ holonomy
is to make a generalisation of the original $G_2$ metric found in
\cite{brysal,gibpagpop} on $\R^4\times S^3$.  We take as our starting
point the metric
\be
ds_7^2 = dt^2 + a_i^2\, (\Sigma_i + g_i\, \sigma_i)^2 + b_i^2 \,
\sigma_i^2\,,\label{7met3}
\ee
where $\sigma_i$ and $\Sigma_i$ are left-invariant 1-forms on two
$SU(2)$ group manifolds, $S_\sigma^3$ and $S_\Sigma^3$, and $a_i$,
$b_i$ and $g_i$ are functions of the radial coordinate $t$.  The
principal orbits are therefore $S^3$ bundles over $S^3$, and since the
bundle is topologically trivial, they have the topology $S^3\times
S^3$.  The isometry group in general is $SU(2)_L^\sigma\times
SU(2)_L^\Sigma$, where the two factors denote left-acting $SU(2)$
transformations on $S^3_\sigma$ and $S^3_\Sigma$ respectively.

    The original metric ansatz used in obtaining the solution in
\cite{brysal,gibpagpop} is obtained by taking $a_i=a$, $b_i=b$, and
$g_i=-\ft12$.  Note that in this ``round'' case the isometry group is
enhanced to $SU(2)_L^\sigma\times SU(2)_L^\Sigma\times SU(2)_R^{D}$,
where the third factor denotes the diagonal $SU(2)$ subgroup of the
product of right-acting $SU(2)_R^\sigma$ and $SU(2)_R^\Sigma$
transformations.  There is a natural triality permutation symmetry
$S_3$, generated by the $\Z_2$ ``flop'' described in \cite{amv} that
interchanges the $SU(2)_L^\sigma$ and $SU(2)_L^\Sigma$ factors, and
another $\Z_2$ ``flip'' operation under which $SU(2)$ group elements
$G_\sigma$ and $G_\Sigma$ on the two 3-spheres are transformed
according to $(G_\sigma,G_\Sigma)\longrightarrow (G_\sigma^{-1},
G_\Sigma\, G_\sigma^{-1})$.  These will be discussed in more detail in
section 2.3.

   It is straightforward to obtain the conditions for Ricci-flatness
of the metric (\ref{7met3}).  We find that they can be derived from
the Lagrangian $L=T-V$, together with the constraint $T+V=0$, where
$T=\ft12 g_{ij}\, (d\a^i/d\eta)\, (d\a^j/d\eta)$, and the $9\times 9$
metric on the ``non-linear sigma model'' for the functions in the
ansatz is given by
\be
g_{ij} =\hbox{block-diag}(2-2\delta_{ab},
 -a_1^2\, b_1^{-2}, -a_2^2\, b_2^{-2}, -a_3^2\, b_3^{-2})\,,
\label{smet1}
\ee
and $V=(\prod_i a_i\, b_i)^2\, (U_1+U_2 +U_3)$, with
\bea
U_1 &=& -\fft1{a_1^2} + \fft{a_1^2}{2a_2^2\, a_3^2} - \fft{1}{b_1^2}
  +\fft{b_1^2}{2b_2^2\, b_3^2} \nn\\
  && +g_1^2\, \Big(\fft{a_1^2}{2b_2^2\, b_3^2} + \fft{a_2^2}{2 b_1^2\,
a_3^2} + \fft{a_3^2}{2b_1^2\, a_2^2} \Big) + \fft{a_1^2}{2b_2^2\,
b_3^2} \, ( g_2^2 \, g_3^2 + 2 g_1\, g_2\, g_3)\,.
\eea
The functions $U_2$ and $U_3$ are related to $U_1$ by cyclic
permutation of the indices 1, 2 and 3.  The 9 functions $\a^i$ are
defined by
\be
\a^i= \log a_i\qquad \a^{i+3}=\log b_i\,,\qquad \a^{i+6}= g_i\,,\qquad
i=1,2,3\,,
\ee
and the indices $a$ and $b$ in (\ref{smet1}) range over the $6\times
6$ block with $1\le a\le 6$ and $1\le b\le 6$.  The radial coordinate
$\eta$ is related to $t$ by $dt= (\prod_i a_i\, b_i)\, d\eta$.

\subsection{Specialisation to the ``round'' metric, and 
its general solution}

\subsubsection{Superpotential and first-order equations}

   There does not appear to exist a superpotential and associated
first-order equations for the 9-function ansatz (\ref{7met3}).
However, we have found a superpotential for the case where we take all
three directions equal, so that just three functions remain.  It is
convenient to write these as
\be
a_i =a\,,\qquad b_i = b\,,\qquad g_i = g-\ft12\,,\label{3fun}
\ee
and so the ansatz (\ref{7met3}) now reduces to
\be
ds_7^2= dt^2 + a^2\, [\Sigma_i+(g-\ft12)\, \sigma_i]^2 + b^2\,
\sigma_i^2\,.\label{7met4}
\ee
Since in this case both the base and the fibre are round three
spheres, giving th enhanced $SU(2)^3$ symmetry discussed in section
2.1, we shall refer to this as the ``round $G_2$ manifold.''

  The Lagrangian $L=T-V$ describing the Ricci-flat equations for this
is given by $T=\ft12 g_{ij}\, (d{\a^i}/d\eta)\, (d{\a^j}/d\eta)$,
where $\a^1\equiv \log a$, $\a^2\equiv \log b$ and $\a^3\equiv g$,
$\eta$ is defined by $dt= a^3\,
c^3\, d\eta$, and
\be
g_{ij} = \pmatrix{12 & 18 & 0\cr
                  18 & 12 & 0 \cr
                  0  & 0 & -\fft{3a^2}{b^2}}\,.
\ee
The Lagrangian is of a ``non-linear sigma model'' type, where the
metric $g_{ij}$ depends on the functions in the system.  We find that
there is a superpotential for this system, such that $V=-\ft12
g^{ij}\, (\del W/\del \a^i)\, (\del W/\del \a^j)$, where $W$ is given
by
\be
W =\fft34 a^2\, b\, K\,,
\ee
and
\be
K\equiv \sqrt{[a^2\, (1-2g)^2 + 4 b^2]\, [a^2\, (1+2g)^2 + 4 b^2]}\,.
\ee
The associated first-order equations, expressed in terms of the
original radial variable $t$, are then
\bea
\dot a &=& \fft{16 b^4 - (1-4g^2)^2\, a^4}{8b^2\, K}\,,\qquad
\dot b = \fft{a\, [a^2\, (1-4g^2)^2 + 4 (1+4g^2)\, b^2]}{4b\,
K}\,,\nn\\
\dot g &=& \fft{2g\, [a^2\, (1-4g^2) - 4 b^2]}{a\, K}\,.
\label{3funeq}
\eea

\subsubsection{Covariantly-constant spinor and $G_2$ holonomy}

    We now show that the first-order equations (\ref{3funeq}) are the
integrability conditions for the existence of a covariantly-constant
spinor.  In turn, this demonstrates that the metrics (\ref{7met3})
with functions given by (\ref{3fun}) satisfying (\ref{3funeq}) have
the special holonomy $G_2$.

    From the spin connection for the metric, we find that in the
obvious choice of orthonormal basis $e^0=dt$, $e^i= b\, \sigma_i$ and
$e^{\td i} = a\, (\Sigma_i + (g-\ft12)\, \sigma_i)$, and spin frame, a
covariantly-constant spinor $\eta$ satisfying $D\eta\equiv d\eta +
\ft14 \omega_{ab}\, \Gamma_{ab}\, \eta=0$ is independent of the
coordinates of the $S^3\times S^3$ principal orbits, but it does
depend on the radial coordinate $t$.  After some algebra we find that
if the first-order equations (\ref{3funeq}) are satisfied there is
exactly one solution, which is given by
\be
\eta= f_1\,\eta_1 + f_2\,\eta_2\,,
\ee
where
\bea
f_1 &=& \fft{(a(1-2g)-2{\rm i}\, b)(a+2{\rm i}\, b)}{
\sqrt{(a^2(1-2g)^2 + 4b^2)(a^2 +4b^2)}}\,,\nn\\
f_2 &=& \fft{(a(1+2g)-2{\rm i}\,b)
(a+2{\rm i}\, b)}{\sqrt{(a^2(1+2g)^2 + 4b^2)(a^2 + 4b^2)}}\,,
\eea
where $\eta_1$ and $\eta_2$ are constant spinors, satisfying
the constraints
\be
(\Gamma_1-{\rm i}\, \Gamma_4)\,\eta_2=0\,,\qquad
\Gamma_{56}\,\eta_1={\rm i}\, \Gamma_{01}\, \eta_2\,,\qquad
(\Gamma_{26}-\Gamma_{35})\,\eta_1=2\Gamma_{01}\,
\eta_2\,.\label{roundkscon}
\ee
Here the explicit index values 1, 2 and 3 refer to the vielbein
components $e^i$ in the $S^3$ base manifold, while 4, 5 and 6 refer to
the vielbein components $e^{\td i}$ in the $S^3$ fibres.

\subsubsection{The general solution to the first-order equations}

   We can obtain the general solution to the first-order equations
(\ref{3funeq}) as follows.  Defining a new radial coordinate $w$ by
$dw=a\, K^{-1}\, dt$, and new variables $\a$, $\beta$ and $\gamma$ by
$a^2=\a$, $b^2=\beta$ and $g^2=\ft14 \gamma$, the first-order
equations become
\bea
\fft{d\a}{dw} &=& 4 \beta - \fft{\a^2\, (1-\gamma)^2}{4\beta}\,,\nn\\
\fft{d\beta}{dw} &=& \ft12 \a\, (1-\gamma)^2 + 2\beta\,
(1+\gamma)\,,\\
\fft{d\gamma}{dw} &=& 4\gamma\, (1-\gamma) - \fft{16 \beta\,
\gamma}{\a}\,.\nn
\eea
which, after the further redefinition $\gamma=\psi^{-4}$, becomes
\be
\fft{d^2\psi}{dw^2} + 4 \psi^{-3} - 4\psi=0\,.
\ee
From this we obtain $(d\psi/dw)^2= \td k + 4\psi^2 + 4\psi^{-2}$, and hence,
after writing $e^{4w}= \td k\, \rho$ and then $\td k= 8/\lambda$, we
arrive at the general solution
\be
a^2=  F^{-1/3}\, Y\,,\qquad b^2 = \ft14 F^{2/3}\, Y^{-1}\,,
\qquad g= m\, \lambda\, \rho\, Y^{-1}\,,\label{abgsol}
\ee
where
\bea
F &\equiv& 3\rho^4 - 8m\, \rho^3  + 6m^2\, (1-\lambda^2)\,\rho^2 - m^4\,
(1-\lambda^2)^2\,,\nn\\
Y &\equiv&\rho^2 -2m\, \rho + m^2\, (1  -\lambda^2)\,,
\label{fydef}
\eea
(Note that $dF/d\rho=12\rho\, Y$.)  Thus the metric in this general
solution is given by
\be
ds_7^2 = F^{-1/3}\, d\rho^2 + F^{-1/3}\, Y\,
(\Sigma_i + (g-\ft 12)\, \sigma_i)^2 + \ft14 F^{2/3}\,
Y^{-1}\, \sigma_i^2\,.\label{genmet}
\ee
The metric is asymptotically conical (AC), with base the non-product
Einstein metric on $S^3\times S^3$.  Note that if $m=0$, the $\lambda$
dependence of the solution drops out, and the metric becomes simply
the Ricci-flat cone over the $S^3\times S^3$ base.

   When $|\rho|$ is large, the functions $F$ and $Y$ are both large and
positive, with $F\sim 3 \rho^4$, $Y\sim \rho^2$ and $g\sim 0$.  Letting
$\rho =r^3$, we see that at large $r$, after dropping an overall
constant scale factor $12 (3)^{1/3}$, we shall have
\be
ds_7^2 \sim d r^2 + \ft19 r^2 \,
(\Sigma_i-\ft12 \sigma_i)^2 + \ft1{12} r^2\, \sigma_i^2\,.
\ee
Thus in this region the metric is asymptotic to the cone over
$S^3\times S^3$.

  As $|\rho|$ reduces from infinity, the metric remains regular until
the first singularity in the metric functions is encountered.  Let us
first check the function $Y$.  This has zeros at $\rho=m\,
(1\pm\lambda)$.  Substituting these values into $F$, we see that it is
given by $F=-4(1\pm\lambda)^2\, \lambda^2\, m^4$.  Thus leaving aside
the special cases $\lambda=0$ or $\lambda=\pm 1$ for now, we see from
the fact that $F$ is negative here that the first zero of $F$ must
have occurred {\it before} $Y$ reached its first zero.  Thus when
$\lambda\ne \pm1$ or 0, it is the first zero of $F$, and not $Y$, that
defines the inner endpoint $\rho=\rho_0$ of the range of the radial
variable.  This occurs at $F(\rho_0)=0$, at which point $Y(\rho_0)$
will still be positive; we shall define $Y_0\equiv Y(\rho_0)>0$.

    Near to $\rho=\rho_0$, we shall have $F(\rho) \approx
(\rho-\rho_0)\, F'(\rho_0)$, which means $F(\rho)\approx
12(\rho-\rho_0)\, Y_0$.  Letting $12(\rho-\rho_0)\,
Y(\rho_0)=r^{6/5}$, we see that near $\rho=\rho_0$ the metric takes
the form
\be
ds_7^2 \sim \fft{1}{25 Y_0^2}\, \Big[ d r^2 + 100 Y_0^3\,
r^{-2/5}\, [\Sigma_i+(b-\ft12)\, \sigma_i]^2 + 25 Y_0\,
r^{4/5}\, \sigma_i^2\Big]\,,
\ee
with $b=8\rho_0/Y_0$. The metric is clearly singular at $r=0$ (\ie at
$\rho=\rho_0$).

    Having shown that the metric is singular for general values of the 
non-trivial parameter $\lambda$, we now consider the three special
values, $\lambda=0$ and $\lambda=\pm1$, for which were excluded from
the general analysis.

\subsubsection{The non-singular solutions}

   There are three values of the non-trivial integration constant
$\lambda$ for which we find that the generically singular $G_2$ metric
(\ref{genmet}) is regular, namely $\lambda=0$ and $\lambda=\pm1$.  We
shall consider these one by one.
\medskip

\noindent
$\bullet \quad \underline{\lambda=0}$: In this case, we have
\be
F=(3\rho+m)(\rho-m)^3\,,\qquad
Y=(\rho-m)^2\,,\qquad g=0\,.\label{l0sol}
\ee
If we define a new radial coordinate
$r=\sqrt{3}\, (3\rho+m)^{1/3}$ and scale parameter $\ell^3=4\sqrt3\,
m$, the metric (\ref{genmet}) then becomes
\be
ds_7^2 = \fft{dr^2}{\Big(1-\fft{\ell^3}{r^3}\Big)} + \ft19\, r^2
\, \Big(1-\fft{\ell^3}{r^3}\Big)\, (\Sigma_i -\ft12 \sigma_i)^2 +
\ft1{12}\, r^2\, \sigma_i^2\,.\label{oldmet}
\ee
This is the previously-known complete metric of $G_2$ holonomy on the
$\R^4$ bundle over $S^3$, as found in \cite{brysal,gibpagpop}.

\medskip

\noindent
$\bullet \quad \underline{\lambda= -1}$:  Now we have
\be
F= \rho^3\, (3\rho-8m)\,,\qquad Y= \rho\, (\rho-2m)\,,\qquad
g= -\fft{m}{(\rho-2m)}\,.
\ee
After making the redefinitions $r=\sqrt3\, (3\rho-8m)^{1/3}$ and
$\ell^3 = -24\sqrt3\, m$, the metric (\ref{genmet}) becomes
\be
ds_7^2 = \fft{dr^2}{\Big(1-\fft{\ell^3}{r^3}\Big)} +
    \ft19 r^2\, \Big(1-\fft{\ell^3}{4r^3}\Big)\, [\Sigma_i +(g-\ft12)\,
\sigma_i]^2 + \ft1{12} r^2\, \fft{\Big(1-\fft{\ell^3}{r^3}\Big)}{
 \Big(1-\fft{\ell^3}{4r^3}\Big)}\, \sigma_i^2\,,\label{newold1}
\ee
with
\be
g = \fft{3\ell^3}{
              8\Big(1-\fft{\ell^3}{4r^3}\Big)}\,.\label{newoldb1}
\ee
This metric is clearly regular, and indeed it can easily be seen to be
equivalent to the original metric of \cite{brysal,gibpagpop}.  This
can be done by first ``recompleting the square'' so that the terms in
the metric are instead organised with the structure
$x_1\, (\sigma_i + y\, \Sigma_i)^2 + x_2\, \Sigma_i^2$, and then
making the transformation
\be
\sigma_i\longrightarrow \Sigma_i\,,\qquad \Sigma_i\longrightarrow
\sigma_i\,.
\ee
After doing this, the metric becomes precisely (\ref{oldmet}).  Note
that this $\Z_2$ transformation is precisely the ``flop'' described in
\cite{amv}.

\medskip

\noindent
$\bullet \quad \underline{\lambda=+1}$: In this case we have
\be
F= \rho^3\, (3\rho-8m)\,,\qquad Y= \rho\, (\rho-2m)\,,\qquad
g= \fft{m}{(\rho-2m)}\,.
\ee
Under the same transformations $r=\sqrt3\, (3\rho-8m)^{1/3}$ and
$\ell^3 = -24\sqrt3\, m$ as for the case $\lambda=-1$, the metric
(\ref{genmet}) becomes
\be
ds_7^2 = \fft{dr^2}{\Big(1-\fft{\ell^3}{r^3}\Big)} +
    \ft19 r^2\, \Big(1-\fft{\ell^3}{4r^3}\Big)\, [\Sigma_i +(g-\ft12)\,
\sigma_i]^2 + \ft1{12} r^2\, \fft{\Big(1-\fft{\ell^3}{r^3}\Big)}{
 \Big(1-\fft{\ell^3}{4r^3}\Big)}\, \sigma_i^2\,,\label{newold2}
\ee
with
\be
g = -\fft{3\ell^3}{
              8\Big(1-\fft{\ell^3}{4r^3}\Big)}\,.\label{newoldb2}
\ee
This differs from $\lambda=-1$ case only in the sign of $g$, and the
metric is again regular.  However, if one repeats the previous
procedure of recompleting the square and interchanging $\sigma_i$ and
$\Sigma_i$, one now finds that (\ref{newold2}) simply maps into
itself.  It is, nevertheless, equivalent to the original metric
(\ref{oldmet}), but the required coordinate transformation that
explicitly implements this is more complicated.  This will be
discussed in the next subsection.

\subsection{$S_3$ transformations of the round $G_2$ metrics}

  In this subsection, we shall show that there is an $S_3$ triality
transformation of the metric (\ref{genmet}) that maps it back into its
original isotropic form, with transformed values of its
parameters.\footnote{This triality symmetry among the regular $G_2$
holonomy metrics was discussed in detail in \cite{aw}. M.C.~would
like to thank E.~Witten for sharing with her the draft of \cite{aw}
prior to the publication. } The three permutation transformations can
be generated by two types of operation.  The first is a generalisation
of the one we already used in order to transform $\lambda=-1$ regular
metric into the $\lambda=0$ metric, which amounts to an exchange of
the base and the fibre 3-spheres.  The second is a more complicated
one, involving transformations of the coordinates on the 3-spheres.
We shall discuss these in succession.
\medskip

\noindent $\bullet\quad$\underline{The interchange of the 3-spheres}

    This transformation follows from the ``recompletion of squares''
identity
\be
x_1\, (\Sigma_i + y\, \sigma_i)^2 + x_2\, \sigma_i^2 =
\hat x_1\, (\sigma_i + \hat y\, \Sigma_i)^2 + \hat x_2\, \Sigma_i^2
\,,\label{sS}
\ee
where
\be
\hat x_1 = x_1\, y^2 + x_2\,,\qquad \hat x_2 = \fft{x_1\, x_2}{x_1\,
y^2 + x_2}\,,\qquad \hat y= \fft{x_1\, y}{x_1\, y^2 + x_2}
\,.\label{attr}
\ee
Since after exchanging $\sigma_i$ and $\Sigma_i$ this leaves the
metric (\ref{genmet}) in the same class of ``round'' metrics as
(\ref{7met3}), (\ref{3fun}), and since (\ref{genmet}) is the general
solution of the first-order equations, it follows that the transformed
metric must be contained within the general solutions (\ref{genmet}),
after appropriate transformations of the parameters.\footnote{There
are really three parameters in the general solutions, namely the
non-trivial parameter $\lambda$, the scale parameter $m$, and a
trivial constant shift of the radial coordinate that we have not made
explicit in (\ref{genmet}).  Our statement about the mapping of
solutions into solutions must be understood to include the need to
perform such a constant shift.}

   It is easily seen from (\ref{attr}) that the effect of the
$\sigma_i\leftrightarrow \Sigma_i$ flop automorphism is to map the solutions
as follows:
\bea
m &\longrightarrow & \hat m= \ft12 m\, (3\lambda -1)\,,\nn\\
\lambda &\longrightarrow & \hat \lambda =
\fft{\lambda+1}{3\lambda-1}\,,\\
r &\longrightarrow & \hat r = r + m\, (\lambda-1)\,.\nn
\eea
Clearly this is an involution, giving a $Z_2$ transformation on the
solutions.

    A special case of the above is to take $\lambda=-1$, which maps to
$\hat \lambda=0$.  This is the transformation that we used in the
previous subsection to show that (\ref{newold1}) is equivalent to the
original $G_2$ metric (\ref{oldmet}).  On the other hand if we apply
it to the case $\lambda=1$, we see that this is a fixed point of the
transformation, which was already observed in our discussion of the
metric (\ref{newold2}).

\medskip
\noindent$\bullet\quad$ \underline{$SU(2)$ transformations}

    Let $\tau_i$ be the Pauli matrices, and let $U$ and $V$ be
$2\times 2$ matrices for two $SU(2)$ groups.  Then we can define
the left-invariant 1-forms $\sigma_i$ and $\Sigma_i$ as follows:
\be
U^{-1}\, dU = \ft{\im}{2}\, \tau_i\, \sigma_i\,,\qquad
V^{-1}\, dV = \ft{\im}{2}\, \tau_i\, \Sigma_i\,.
\ee

    Now, make a coordinate transformation to new $SU(2)$ matrices
$\wtd U$ and $\wtd V$, related to $U$ and $V$ as follows:
\be
U= \wtd U^{-1}\,,\qquad V = \wtd V\, \wtd U^{-1}\,.\label{su2trans}
\ee
(This is the $(G_\sigma, G_\Sigma)\longrightarrow (G_\sigma^{-1},
G_\Sigma\, G_\sigma^{-1})$ automorphism of the round metrics that we
mentioned in section 2.1.)  It follows that we shall have
\be
U^{-1}\, dU = -\wtd U\, (\wtd U^{-1}\, d\wtd U)\, \wtd U^{-1}\,,\quad
V^{-1}\, dV = \wtd U\, (\wtd V^{-1}\, d\wtd V - \wtd U^{-1}\, d\wtd
U)\, \wtd U^{-1}\,.
\ee
We also define ``tilded'' left-invariant 1-forms
\be
\wtd U^{-1}\, d\wtd U = \ft{\im}{2}\, \tau_i\, \td\sigma_i\,,\qquad
\wtd V^{-1}\, d\wtd V = \ft{\im}{2}\, \tau_i\, \wtd\Sigma_i\,.
\ee

    Since we can write
\be
\sigma_i^2 = -2\tr(\ft{\im}{2}\, \tau_i\, \sigma_i)^2 =
 -2\tr( U^{-1}\, dU)^2\,,
\ee
and so on, it now follows that
\be
x_1\, [\Sigma_i + (g-\ft12)\, \sigma_i]^2 + x_2\, \sigma_i^2
= x_1\, [\wtd\Sigma_i +(-g-\ft12)\, \td\sigma_i]^2 + x_2\,
\td\sigma_i^2\,,\label{su2trans2}
\ee
In other words, the net effect of the transformation (\ref{su2trans})
is simply to send $g\longrightarrow -g$.  Expressed as a
transformation of the parameters in the general solution
(\ref{genmet}), this therefore just amounts to
\be
\lambda \longrightarrow \td\lambda =-\lambda\,,
\ee
while leaving the other parameters unchanged.  Thus we see that the
$S_3$ permutation group acting on the parameter $\lambda$ is generated by
$\lambda\rightarrow (\lambda+1)/(3\lambda-1)$ and $\lambda\rightarrow
 -\lambda$.  The fundamental domain of $\lambda$ is therefore given by
\be
0\le \lambda\le \ft13\,.
\ee
For a generic value of $m$, the solutions form triplets under the
$S_3$ group.  When $m=0$, corresponding to the Ricci-flat
cone over $S^3\times S^3$, the triplet degenerates to a singlet.

    A particular application of this transformation is to the metric
(\ref{newold2}), which corresponds to $\lambda=+1$.  It is therefore
mapped into the previous example (\ref{newold1}), which corresponds to
$\lambda=-1$.  Indeed, we may note that as solutions of (\ref{genmet})
these metrics differ only in the sign of $g$.  This, therefore,
provides the additional transformation that is needed in order to show
that (\ref{newold2}) is again equivalent to the original $G_2$ metric
(\ref{oldmet}).

   As another application of the general result (\ref{su2trans}), we
may note that
\be
(\sigma_i-\Sigma_i)^2 = \wtd\Sigma_i^2\,,\qquad
(\sigma_i+\Sigma_i)^2 = 4 (\td\sigma_i-\ft12 \wtd\Sigma_i)^2\,.
\ee
This enables us to rewrite the original $G_2$ metric (\ref{oldmet}) in
a form that is manifestly invariant under the interchange of
$\sigma_i$ and $\Sigma_i$, namely
\be
ds_7^2 = \fft{dr^2}{\Big( 1-\fft{\ell^3}{r^3}\Big)}
+ \ft1{12} r^2\, (\sigma_i-\Sigma_i)^2
    + \ft1{36}\, r^2 \, \Big(1-\fft{\ell^3}{r^3}\Big)\,
      (\sigma_i+\Sigma_i)^2\,.\label{z2form}
\ee
As we shall see below, this basis is closely related to the one used
in \cite{cglp1} for constructing the metric on the cotangent bundle of
$S^3$.

\subsection{$S^1$ reduction to type IIA}

       Since the $G_2$ manifold is Ricci-flat with special holonomy,
it provides a natural compactification (reduction) space for M-theory.
The vacuum solution is just a metric product of four-dimensional
Minkowski space-time and that of the $G_2$ manifold.  Because of the
discrete triality automorphism of the round G$_2$ manifold, it follows
that there are three different ways of reducing it on $S^1$, so that
the principal orbits of the lower dimensional manifold are the
$T^{1,1}=SO(4)/SO(2)$ coset space.\footnote{Some further details of
the relation to $T^{1,1}$ and the six-dimensional Stenzel manifold
$T^*S^3$ are discussed in section 3.}  In order to do so, we note that
the $\Sigma_i$ can be written in terms of Euler angles as
\be
\Sigma_1=\cos\psi\, d\theta + \sin\psi\, \sin\theta\, d\varphi\,,\quad
\Sigma_2=-\sin\psi\, d\theta + \cos\psi\, \sin\theta\, d\varphi\,,
\quad
\Sigma_3 = d\psi + \cos\theta\, d\varphi\,.
\ee
One can then verify that \cite{clp0}
\be
\sum_i[\Sigma_i + (g-\ft12)\, \sigma_i]^2
=[\cos\theta\, d\psi + (g-\ft12)\,\mu^i\, \sigma_i]^2 +
\sum_i (D_g\mu^i)^2\,,
\ee
where
\be
\mu_1=\sin\theta\,\sin\psi\,,\qquad
\mu_2=\sin\theta\,\cos\psi\,,\qquad
\mu_3=\cos\theta\,,
\ee
and $D_g\mu_i=d\mu_i + \epsilon_{ijk}\, (g-\ft12)\,\sigma_j\,\mu_k$.
The $\lambda=0$ solution therefore reduces to become a 
wrapped D6-brane in type IIA:
\bea
ds_{10}^2 &=& e^{-\ft16\phi}\, \Big[dx^\mu\, dx_\mu +
\fft{dr^2}{1-\fft{\ell^3}{r^3}} + \ft19r^2 (1-\fft{\ell^3}{r^3})\,
(D_{g=0}\mu^i)^2 + \ft1{12}r^2\, \sigma_i^2\Big]\,,\nn\\
e^{\ft43\phi} &=& \ft19 r^2(1-\fft{\ell^3}{r^3})\,,\qquad
{\cal A}_\1 = \cos\theta\, d\psi + \ft12\mu_i\,\sigma_i\,.
\eea
The six-dimensional transverse space here has a similar structure to
the deformed conoifold.  The $\lambda=-1$ solution becomes
\bea
ds_{10}^2 &=& e^{-\ft16\phi}\, \Big[dx^\mu\, dx_\mu +
\fft{dr^2}{1-\fft{\ell^3}{r^3}} +
\fft19r^2\, (1-\fft{\ell^3}{4r^3}) (D_g\mu^i)^2 +
\ft1{12} r^2\, \fft{1-\fft{\ell^3}{r^3}}{1-\fft{\ell^3}{4r^3}}\,
\sigma_i^2\Big]\,,\nn\\
e^{\ft43\phi}&=&\ft19 r^2\, (1-\fft{\ell^3}{4r^3})\,,\qquad
{\cal A}_\1 = \cos\theta\, d\psi  - (g-\ft12)\, \mu_i\,\sigma_i
\,,\label{d6case2}
\eea
where $g$ is given by (\ref{newoldb1}).  The six-dimensional
transverse space has a structure similar to the resolved conifold.
These two reductions were
previously obtained in \cite{amv}.  The $\lambda=0$ reduction is
singular from the ten-dimensional point of view, but the $\lambda=-1$
reduction is regular.

        Employing the $S_3$ triality transformations, discussed in the previous
subsections, we can obtain the third case, namely $\lambda=1$.  The
form of the metric is the same as (\ref{d6case2}) but with $g$
replaced by $-g$.  It should be emphasised that from the type IIA point of
view, the choice of sign for $g$ has a non-trivial effect.  The
$\lambda=0$ solution gives the same topology as the deformed conifold,
whilst the two solutions $\lambda=\pm 1$ give the topology of the
resolved conifold.  These two resolutions arise from the fact that
there can be a sign choice for the Fayet-Illiopoulos term for the
field theory arising from type IIA string theory on a resolved
conifold.\footnote{We thank E.~Witten for making this observation.}

\section{Squashed manifolds with $G_2$ holonomy}

    We can make a different truncation of the 9-function ansatz
(\ref{7met3}), in which six functions remain, which again allows us to
contruct a superpotential whose first-order equations yield first
integrals of the Ricci-flat Einstein equations.  A further truncation
in this example, to an ansatz with four remaining functions, gives the
system considered first in \cite{bggg}.\footnote{We are grateful to
A. Brandhuber, J. Gomis, S.S. Gubser and S. Gukov for informing us of
their four-function ansatz prior to publication.  The truncation of
the nine-function ansatz of section 2 to the six-function
ansatz that we consider here was motivated by their results, together
with results in \cite{cglp1} on the construction of Stenzel metrics.}  We
shall take a somewhat different approach here, and motivate the
six-function and four-function ans\"atze from a six-dimensional
viewpoint, based on earlier results in \cite{cglp1} on the Stenzel
construction of Ricci-flat K\"ahler metrics on the cotangent bundle of
$S^{n+1}$.

    We take as our starting point the construction of the Stenzel
metrics on $T^*S^{n+1}$, using the formalism and notation of
\cite{cglp1}.  The metrics are of cohomogeneity one, with principal
orbits that are $SO(n+2)/SO(n)$. Denoting the left-invariant 1-forms
of $SO(n+2)$ by $L_{AB}$, satisfying
\be
dL_{AB} = L_{AC}\wedge L_{CB}\,,\label{son2}
\ee
 and decomposing under the $SO(n)$ subgroup according to
$A=(1,2,i)$, the following ansatz for Stenzel metrics was
made in \cite{cglp1}:
\be
ds^2 = dt^2 + a^2\, (L_{1i})^2 + b^2\, (L_{2i})^2 + c^2\,
(L_{12})^2\,.
\ee
Note that the 1-forms $L_{1i}$, $L_{2i}$ and $L_{12}$ are precisely
the ones that live in the coset $SO(n+2)/SO(n)$.

   Let us specialise now to $n=1$, which gives a construction of the
6-dimensional Stenzel metric, with principal orbits $SO(4)/SO(2)$.  In
this case, we can now give a natural extension of the ansatz to one
for 7-dimensional metrics of cohomogeneity one, where the principal
orbits are $SO(4)$, by appending an additional term $f^2\,
(L_{34})^2$:
\be
ds_7^2 = dt^2 +  4 a^2\, (L_{1i})^2 + 4b^2\, (L_{2i})^2 + 4c^2\,
(L_{12})^2 + 4f^2\, (L_{34})^2\,.\label{7met1}
\ee
(The factors of 4 are introduced for later convenience.)  It is
helpful now to exploit the local isomorphism between $SO(4)$ and
$SU(2)\times SU(2)$, which is made manifest by forming self-dual and
anti-self-dual combinations of the $L_{AB}$.  A convenient choice for
these combinations is
\bea
&&\sigma_1 = L_{42} + L_{31}\,,\qquad
  \sigma_2 = L_{23} + L_{41}\,,\qquad
  \sigma_3 = L_{34} + L_{21}\,,\nn\\
&&
\Sigma_1 = L_{42} - L_{31}\,,\qquad   \Sigma_2 = L_{23} - L_{41}
\,,\qquad \Sigma_3 = L_{34} - L_{21}\,.
\eea
It now follows from (\ref{son2}) that the $\sigma_i$ and $\Sigma_i$
are left-invariant 1-forms for the two $SU(2)$ factors, satisfying
\be
d\sigma_i= -\ft12 \ep_{ijk}\, \sigma_j\wedge \sigma_k\,,\qquad
d\Sigma_i= -\ft12 \ep_{ijk}\, \Sigma_j\wedge \Sigma_k\,.
\ee
In terms of these, the metric ansatz (\ref{7met1}) becomes
\be
ds_7^2 = dt^2 + a^2\, [(\sigma_1-\Sigma_1)^2 + (\sigma_2-\Sigma_2)^2]
+ b^2\, [(\sigma_1+\Sigma_1)^2 +(\sigma_2+\Sigma_2)^2]
+ c^2\, (\sigma_3-\Sigma_3)^2
+  f^2\, (\sigma_3 + \Sigma_3)^2\,.
\label{bgggan}\ee
The principal orbits are $SO(4)$ (or its double cover Spin(4)), but
the isometry group may be larger in special cases.  The form
(\ref{bgggan}) is the same as the ansatz previously introduced in
\cite{bggg}.  Indeed the above derivation of the ansatz (\ref{bgggan})
shows how the ansatz of \cite{bggg} is related to the construction of
the six-dimensional Stenzel metric \cite{dloc,stenzel,cglp1}, which is
the deformed conifold.  In particular, since
$L_{34}=\ft12(\sigma_3+\Sigma_3)$, the reduction to six dimensions is
achieved by omitting the last term in (\ref{bgggan}).  

    A further generalisation of the ansatz clearly suggests itself,
namely to write
\be
ds_7^2 = dt^2 + \ft12 a_{AB}^2\, (L_{AB})^2\,,
\ee
where the functions $a_{AB}$ depend only on $t$.  In
the $SU(2)\times SU(2)$ basis for $SO(4)$, this becomes
\be
ds_7^2 = dt^2 + a_i^2\, (\sigma_i-\Sigma_i)^2 + b_i^2\, (\sigma_i +
\Sigma_i)^2\,.\label{7met2}
\ee
By completing the  square, it is easily seen that
this ansatz is a special case of our original 9-function ansatz
(\ref{7met3}).

   Note that this ansatz is subsumed in the (\ref{7met3}) with the
specialisation
\be
g_i^2 = 1-\fft{b_i^2}{a_i^2}\,.
\ee

   Calculating the curvature, we find that the conditions for
Ricci-flatness can be derived from the Lagrangian $L=T-V$ for
the functions $a_i$ and $b_i$, together with the constraint $T+V=0$.
Defining $\a^i\equiv \log a_i$ and $\a^{i+3}\equiv \log b_i$, the
kinetic energy is given by $T= \ft12 g_{ab}\,(d{\a^a}/d\eta)\,
(d{\a^b}/d\eta)$ with
\be
g_{ab} = 2-2\delta_{ab}\,,
\ee
where the new radial variable $\eta$ is defined by
$dt=\prod_i(a_i\, b_i)\, d\eta$.
We find that the potential energy $V$ can be derived from a
superpotential $W$, so that $V=-\ft12 g^{ab}\, (\del W/\del\a^a)\,
(\del W/\del \a^b)$, with
\bea
2W &=& -a_1\, a_2\, a_3\, (b_1^2 + b_2^2 + b_3^2)
 + b_1\, b_2\, b_3\, (a_1\, b_1  +a_2\, b_2 + a_3\, b_3)  \nn\\
&& + b_1\, b_2\, a_3\, (a_1^2 + a_2^2) +
     b_1\, b_3\, a_2\, (a_1^2 + a_3^2) +
     b_2\, b_3\, a_1\, (a_2^2 + a_3^2)\,.
\eea
(Other sign choices can also arise, but these just correspond to
relabellings of indices, and so they give equivalent results.)

   From the superpotential we can derive the first-order equations
$d{\a^a}/d\eta = g^{ab}\, \del W/\del\a^b$.  Expressed back in terms
of the original radial variable $t$, these are
\bea
\dot a_1 &=& \fft{a_1^2}{4 a_3\, b_2} + \fft{a_1^2}{4 a_2\, b_3}
      - \fft{a_2}{4b_3}  -\fft{a_3}{4b_2} - \fft{b_2}{4 a_3} -
        \fft{b_3}{4a_2}\,,\nn\\
\dot a_2 &=& \fft{a_2^2}{4 a_3\, b_1} + \fft{a_2^2}{4 a_1\, b_3}
      - \fft{a_1}{4b_3}  -\fft{a_3}{4b_1} - \fft{b_1}{4 a_3} -
        \fft{b_3}{4a_1}\,,\nn\\
\dot a_3 &=& \fft{a_3^2}{4 a_2\, b_1} + \fft{a_3^2}{4 a_1\, b_2}
      - \fft{a_1}{4b_2}  -\fft{a_2}{4b_1} - \fft{b_1}{4 a_2} -
        \fft{b_2}{4a_1}\,,\nn\\
\dot b_1 &=& \fft{b_1^2}{4 a_2\, a_3} - \fft{b_1^2}{4 b_2\, b_3}
      - \fft{a_2}{4a_3}  -\fft{a_3}{4a_2} + \fft{b_2}{4 b_3} +
        \fft{b_3}{4b_2}\,,\label{sixfo}\\
\dot b_2 &=& \fft{b_2^2}{4 a_3\, a_1} - \fft{b_2^2}{4 b_3\, b_1}
      - \fft{a_1}{4a_3}  -\fft{a_3}{4a_1} + \fft{b_1}{4 b_3} +
        \fft{b_3}{4b_1}\,,\nn\\
\dot b_3 &=& \fft{b_3^2}{4 a_1\, a_2} - \fft{b_3^2}{4 b_1\, b_2}
      - \fft{a_1}{4a_2}  - \fft{a_2}{4a_1} + \fft{b_1}{4 b_2} +
        \fft{b_2}{4b_1}\,,\nn
\eea

    We may note that a consistent truncation of these equations is to
set $a_2=a_1$ and $b_2=b_1$, leading to the simpler system of
equations
\bea
\dot a_1&=& \fft{a_1^2}{4 a_3\, b_1} - \fft{a_3}{4b_1} - \fft{b_1}{4
a_3} - \fft{b_3}{4 a_1}\,,\nn\\
\dot a_3&=& \fft{a_3^2}{2 a_1\, b_1} - \fft{a_1}{2b_1} - \fft{b_1}{2
a_1} \,,\nn\\
\dot b_1&=& \fft{b_1^2}{4 a_1\, a_3} - \fft{a_1}{4a_3} - \fft{a_3}{4
a_1} + \fft{b_3}{4 b_1}\,,\label{simeq}\\
\dot b_3&=& \fft{b_3^2}{4a_1^2} - \fft{b_3^2}{4 b_1^2}\,.\nn
\eea
These are the first-order equations found in
\cite{bggg}.\footnote{A. Brandhuber, J. Gomis, S.S. Gubser and
S. Gukov have obtained an explicit solution of these equations, which
yields a new complete non-compact metric of $G_2$ holonomy
\cite{bggg}.  We are grateful to them for telling us about their
results prior to publication.}  The solutions give rise to metrics
with a Taub-NUT like squashing of the $S^3$ fibres (as well as a
squashing of the $S^3$ base), with the circle corresponding to
$(\sigma_3+\Sigma_3)$ tending to a constant radius at infinity.  Thus
the metrics are asymptotically locally conical (ALC).

    We can make a further consistent truncation of the first-order
equations (\ref{simeq}), by setting $a_1=-a_3=a$, $b_1=b_3=b$.  They
then become
\be
\dot a = \fft{b}{2a}\,,\qquad \dot b = \fft14 -\fft{b^2}{4 a^2}\,.
\ee
By the standard methods, we can solve these by defining a new
radial coordinate $\rho$ such that $dt=d\rho/b$, leading to $a^2=\rho$
and $b^2 = \ft13 \rho + k\, \rho^{-1/2}$.  After setting $\rho=r^2$,
and $k=-\ell^3/3$,
we get (dropping an  overall factor of $12$ from the front)
\be
ds_7^2 = F^{-1}\, dr^2 + \ft1{12} r^2\, (\sigma_i-\Sigma_i)^2
    + \ft1{36}\, r^2 F\, (\sigma_i+\Sigma_i)^2\,,\label{z2form2}
\ee
where
\be
F \equiv 1 -\fft{\ell^3}{r^3}\,.
\ee
This is in fact the original $G_2$ metric on $\R^4\times S^3$, written
in the $Z_2$-symmetric form that we presented in (\ref{z2form}).

\subsection{Parallel spinor and $G_2$ holonomy}

    It is a straightforward matter to solve the equations $D\eta\equiv
d\eta + \ft12 \omega_{ab}\, \Gamma^{ab}\, \eta=0$ for
covariantly-constant spinors, in the metrics (\ref{7met2}).  We find
that there is precisely one such spinor, if and only is the six metric
functions $a_i$ and $b_i$ satisfy the first-order equations
(\ref{sixfo}) (or the discrete set of other possibilities
corresponding to sign-convention choices).  Thus the first-order
equations (\ref{sixfo}) coming from the superpotential have the
interpretation of being the integrability conditions for the metrics
(\ref{7met2}) to admit a covariantly-constant spinor, and hence for
them to have $G_2$ holonomy.
 
    In the orthonormal frame with $e^0=dt$, $e^i = a_i\,
(\sigma_i-\Sigma_i)$ and $e^{\td i} = b_i\, (\sigma_i + \Sigma_i)$,
with the obvious choice for the spin frame, we find that the
covariantly-constant spinor $\eta$ has constant components, and it
satisfies the constraints
\be
(\Gamma_{16}+\Gamma_{34})\,\eta =0\,,\qquad
(\Gamma_{35}+\Gamma_{26})\,\eta =0\,,\qquad
(\Gamma_{01}-\Gamma_{26})\,\eta =0\,.\label{cccon}
\ee
Here the indices 1, 2 and 3 refer to the vielbein components $e^i$,
while 4, 5 and 6 refer to $e^{\td i}$.  It is easy to verify that, up
to overall scale, the conditions (\ref{cccon}) specify the spinor
uniquely.  Note that $\eta$ can also be expressed as
$\eta=\eta_1+\eta_2$, where $\eta_1$ and $\eta_2$ satisfy the
constraint (\ref{roundkscon}).

\section{Massless M3-branes}

\subsection{M3-branes on round $G_2$ manifolds}

      In \cite{clpmassless}, M3-brane were constructed on the
background of the deformed $G_2$ manifolds, where the 4-form field
strength of the eleven-dimensional supergravity was turned on.
When the 4-form field strength is set to zero, the solution becomes
the product of the Minkowski space-time and the regular $G_2$
manifold.   In particular, we shall begin by considering the example
studied  in
\cite{clpmassless} that is based on the $G_2$ manifold of the $\R^4$
bundle
over $S^3$.  The metric for the associated massless M3-brane is
\cite{clpmassless}
\be
ds_{11}^2 = H^2\, dx^\mu\, dx_\mu + 12 H^4\, Y^{-1}\,  dr^2
           + \ft43 r^2\, H^{-2}\, Y\, (\sigma_i -\ft12 \Sigma_i)^2
          + r^2\, H^{-2}\,
\Sigma_i^2\,,\label{mets3s3sol}
\ee
where
\be
Y\equiv 1-\fft{\ell^3}{r^3} \,,\qquad
H = \Big( 1 -\fft{c^2}{r^{12}\, Y^3}\Big)^{-1/6}\,.\label{yhsol}
\ee

    We can now perform the two $Z_2$
transformations given in section 2.3, to
obtain the previous M3-brane solution (\ref{mets3s3sol})
written in the other two forms.
Firstly, if we apply the $\sigma_i\leftrightarrow \Sigma_i$
transformation as in (\ref{sS}), then we find that (\ref{mets3s3sol})
becomes
\be
ds_{11}^2 = H^2\, dx^\mu\, dx_\mu + 12 H^4\,
Y^{-1}\, dr^2 + \ft43 r^2\, H^{-2}\,
Z\, (\Sigma_i + v\, \sigma_i)^2 + r^2\,
H^{-2}\, \fft{Y}{Z} \, \sigma_i^2\,,\label{mets3s3sol2}
\ee
where $H$ and $Y$ are given in (\ref{yhsol}) and
\be
Z\equiv 1-\fft{\ell^3}{4r^3}\,,\qquad v= -\fft{Y}{2Z}\,.
\ee

   Instead, we can apply to (\ref{mets3s3sol}) the $SU(2)$
transformation as in (\ref{su2trans}).  This gives the M3-brane in the
form
\be
ds_{11}^2 = H^2\, dx^\mu\, dx_\mu + 12 H^4\, Y^{-1}\, dr^2 + \ft13
r^2\, H^{-2}\, Y\, (\sigma_i+\Sigma_i)^2 + r^2\, H^{-2}\,
(\sigma_i-\Sigma_i)^2\,.\label{mets3s3sol3}
\ee
Indeed, (\ref{mets3s3sol2}) and (\ref{mets3s3sol3}) are equivalent,
however they lead to very different solutions after reduction to type
IIA theory on the circle corresponding to $(\sigma_3+\Sigma_3)$.

        Since the regularity of the $G_2$ background in the above
solutions is in any case lost when the M3-brane is introduced by
turning on the 4-form field strength \cite{clpmassless}) , it is not
unreasonable to construct the more general M3-branes on deformations of
the round $G_2$ manifolds of section 2.2.  In analogy with
\cite{clpmassless}, the metric ansatz will be given by
\be
ds_{11}^2 = H^2\, dx^\mu\, dx_\mu + d\rho^2 + a^2\, h_i^2 + b^2\,
\sigma_i^2\,,
\ee
where $h_i=\Sigma_i + (g-\ft12)\, \sigma_i$.   The ansatz for the
4-form field strength, and its resulting dual, are given by
\bea
F_\4&=& f_1\, h_i\wedge h_j\wedge\sigma_i\wedge \sigma_j +
        f_2\, d\rho\wedge h_1\wedge h_2\wedge h_3 +
        \ft12 f_3\, \epsilon_{ijk}\, d\rho \wedge h_i\wedge\sigma_j
                                     \wedge \sigma_k\nn\\
&&+ \ft12 f_4 \epsilon_{ijk}\, d\rho\wedge h_i\wedge h_j\wedge
\sigma_k\,,\nn\\
H^{-4}{*F_\4}&=&\fft{2f_1}{a\,b}\, d^4x\wedge d\rho\wedge h_i\wedge
\sigma_i + \fft{2f_2\, b^3}{a^3}\, d^4x\wedge \sigma_1\wedge
\sigma_2\wedge \sigma_3 \nn\\
&&+ \fft{f_3\,a}{2b}\, \epsilon_{ijk}\, d^4x\wedge h_i\wedge h_j\wedge
\sigma_k
 -\fft{f_4\, b}{2a}\, \epsilon_{ijk}\, d^4x\, h_i\wedge \sigma_j
\wedge \sigma_k\,.\label{f4sf4}
\eea
Thus we have
\bea
F^2_{00}&=&6(\fft{f_2^2}{a^6} + \fft{3 f_3^2}{a^2\,b^4} +
\fft{3 f_4^2}{a^4\,b^2})\,,\nn\\
F^2_{11}&=&F^2_{22}=F^2_{33} =
6(\fft{8f_1^2}{a^4\,b^4} + \fft{2 f_3^2}{a^2\,b^4} +
      \fft{f_4^2}{a^4\,b^2})\,,\nn\\
F^2_{44}&=&F^2_{55}=F^2_{66} =
6 (\fft{8f_1^2}{a^4\,b^4} + \fft{f_2^2}{a^6} +
   \fft{f_3^2}{a^2\,b^4} + \fft{2f_4^2}{a^4\,b^2})\,,\nn\\
F^2_{14}&=&F^2_{25}=F^2_{36}=
\fft{12 f_3\, f_4}{a^3\,b^3} + \fft{6 f_2\, f_4}{a^5\,b}
\,.
\eea
The Bianchi identity and the equation motion for $F_4$ implies
\bea
dF_\4=0:&&2f_1' + (g^2-\ft14)\,f_2 + f_3 +2g\, f_4=0\,,\nn\\
d{*F_\4}=0:&& f_4=2a^2\, b^{-2}\, g\, f_3\,,\nn\\
   &&   b^3\, a^{-3}\, f_2 -3(g^2-\ft14)\,a\, b^{-1}\, f_3
=\lambda\, H^{-4}\,,\nn\\
&&(a\, b^{-1}\, H^4\, f_3)' + 2 a^{-1}\, b^{-1}\,
H^4\, f_1=0\,.
\eea
We find that the existence of a superpotential of such a system seems
to require that the constant $\lambda=0$.  We shall set this constant
to zero from now on.  Thus we find that the 4-form ansatz is
determined by one function only, namely $f_3$.  Thus, the total number
of functions in our ansatz is 5, consisting of $a, b, g, H$ and $f_3$.
After some algebra, we find that the Einstein equations and that for
$f_3$ can be obtained from the Lagrangian $L=T-V$, where the kinetic
term $T$ can be expressed as $T=\ft12 g_{ij}\, \dot \alpha^i\, \dot
\alpha^j$, where $\alpha^i=(\log a, \log b, \log H, g, f_3)$, and a
dot denotes a derivative with respect to $\eta$ defined by $dt=a^3\,
b^3\, H^4\, d\eta$.  The metric $g_{ij}$ is somewhat complicated, but
its inverse is simpler, which we present here
\be
g^{ij}=\fft{1}{18}\pmatrix{-2 & 1 & 1 & 0 & -f_3\cr
                          1 &-2& 1 & 0 &-7f_3 \cr
                          1 & 1 & -\ft45 & 0 & 5f_3\cr
                          0&0&0& -\fft{6b^2}{a^2} & 0 \cr
      -f_3&-7f_3&5f_3 &0 & 2(3b^4-13f_3^2)}\,.
\ee
The potential can be expressed in terms of a superpotential $W$,
namely $V=-\ft12 g^{ij}\, \del_i W\, \del_j W$, where $W$ is given by
\be
W=\ft34 a^2\, b\, H^4\, K_1\, K_2\,,
\ee
where
\be
K_1=\sqrt{[a^2(1-2g)^2+4b^2][a^2(1+2g)^2+4b^2]}\,,\qquad
K_2=\sqrt{1-b^{-4}\, f_3^2}\,.
\ee

       The first-order equations can now be straightforwardly obtained,
and they are given by
\bea
&&a'=\fft{16b^8 -a^4\, (1-4g^2)^2\, (b^4-2f_3^2) +
     8a^2\, b^2\, (1+4g^2)\,f_3^2}{8 b^6\, K_1\, K_2}\,,\nn\\
&&b'=\fft{a^4\,(1-4g^2)^2\,(2b^4-f_3^2) + 8a^2\, b^6\,
(1+4g^2) + 16b^4\, f_3^2}{8 a\, b^5\,  K_1\, K_2}\,,\nn\\
&&g'=-\fft{2g\,(4b^2 - a^2\, (1-4g^2))K_2}{a\, K_1}\,,\qquad
H'=-\fft{f_3^2\,K_1}{8 a\, b^6\, K_2}\,,\label{roundfo}\\
&&f_3'=\fft{f_3\, (16 b^4\,(-3b^4+5f_3^2) + a^4\, (1-4g^2)^2\,
(b^4+f_3^2) -8 a^2\, b^2\, (1+4g^2)(b^4-3f_3^2))}{8a\, b^6\,
K_1\, K_2}\,.\nn
\eea
To solve them we begin by defining new hatted variables
\be
\hat a \equiv a\, H\,,\qquad \hat b \equiv b\, H\,,\qquad
\hat f_3 \equiv f_3\, H^2\,,\qquad
\ee
and a new radial coordinate $x$ by $dx= H\, K_2\, d\rho$.  In terms of
these, we find that the first-order equations (\ref{roundfo}) become
\bea
\fft{d\hat a}{dx}  &=& \fft{16 \hat b^4 -
            (1-4g^2)^2\, \hat a^4}{8\hat b^2\, \hat K_1}\,,\qquad
\fft{d\hat b}{dx} = \fft{\hat a\, [\hat a^2\, (1-4g^2)^2 +
                    4 (1+4g^2)\, \hat b^2]}{4\hat b\, \hat K_1}\,,\nn\\
\fft{d g}{dx} &=& \fft{2g\, [\hat a^2\, (1-4g^2)
               - 4 \hat b^2]}{\hat a\, \hat K_1}\,,\qquad
\fft{1}{H}\, \fft{dH}{dx} = - \fft{ \hat f_3^2\, \hat K_1}{8 \hat a\,
\hat b^6\, \hat K_2^2}\,,\nn\\
\fft{d \hat f_3}{dx} &=& - \fft{\hat f_3 [48\hat b^4 - \hat a^4\,
(1-4g^2)^2 + 8\hat a^2\, \hat b^2\, (1+4g^2)]}{8 \hat a\, \hat b^2\,
\hat K_1}\,,\label{hatroundfo}
\eea
where
\be
\hat K_1\equiv \sqrt{[\hat a^2(1-2g)^2+4\hat b^2]
               [\hat a^2(1+2g)^2+4\hat b^2]}\,.
\ee
We also find that
\be
\hat a^3\, \hat b\, \hat f_3 = \kappa\,,\label{const}
\ee
where $\kappa$ is a constant of integration.

    The general solution of these equations can be obtained as
follows.  To begin with, we note that the first three equations in
(\ref{hatroundfo}) are precisely the same as (\ref{3funeq}), in terms
of the new variables.  Thus the solution for $\hat a$, $\hat b$ and
$g$ as functions of $x$ will be identical that for $a$, $b$ and $g$ as
functions of $t$ for the Ricci-flat $G_2$ holonomy 7-metrics
(\ref{7met4}).\footnote{This phenomenon, where functionally-rescaled
metric components in the 7-dimensional space transverse to the
massless M3-brane turn out to satisfy the same first-order equations
as the metric components of the original Ricci-flat 7-metrics, was
seen also in the original examples obtained in \cite{clpmassless}.}
Thus in terms of a new radial variable $r$, defined by $dx =
F^{-1/6}\, dr$, we obtain from (\ref{abgsol}),
\be
\hat a^2=  F^{-1/3}\, Y\,,\qquad \hat b^2 = \ft14 F^{2/3}\, Y^{-1}\,,
\qquad g= m\, \lambda\, r\, Y^{-1}\,,\label{abgsol2}
\ee
where
\bea
F &\equiv& 3r^4 - 8m\, r^3  + 6m^2\, (1-\lambda^2)\,r^2 - m^4\,
(1-\lambda^2)^2\,,\nn\\
Y &\equiv& r^2 -2m\, r + m^2\, (1  -\lambda^2)\,.
\label{fydef2}
\eea
From (\ref{const}) we then obtain
\be
\hat f_3 = 2\kappa\, F^{1/6}\, Y^{-1}\,.
\ee
Finally, solving the equation for $H$ in (\ref{hatroundfo}), and
fixing a constant of integration without loss of generality, we obtain
\be
H = (1 - \fft{64\,\kappa^2}{F})^{-1/6}\,.
\ee
In general the new Ricci-flat $G_2$ manifolds found in section 2 are
singular where $F=0$.  Here, we see that when the 4-form field
strength is turned on, the associated M3-brane metric becomes singular
before this point is reached, namely at the value ofr $r$ for which
$F=64\kappa^2$ (and hence $H=0$).  Again, the function $H$ falls off
too rapidly as a function of the proper radial distance for the
M3-brane to have non-zero charge and ADM mass.

     It is worth noting that in this case, as in the previous M3-brane
examples in \cite{clpmassless}, the integration constant $\kappa$
appears in the 4-form field strength linearly, but in the metric
quardratically.  It follows that also in this case the metric remains
real if $\kappa$ is replaced by ${\rm i}\, \kappa$.  As discussed in
\cite{clpmassless}, the replacement $\kappa\longrightarrow {\rm i}\,
\kappa$ can be viewed as the replacement of a real 4-form by a
Hodge-dual 3-form that is again real, in the positive-definite
7-dimensional space obtained by Kaluza-Klein reduction on
4-dimensional world-volume of the M3-brane.  In this dual description,
the 3-form in $D=7$ can be viewed as arising from the world-volume
dimensional reduction of an NS-NS 2-brane configuration in ten
dimensions.  Thus the replacement $\kappa\longrightarrow {\rm i}\,
\kappa$ induces a transition between the
M3-brane and an NS-NS 2-brane configuration.

\subsection{M3-branes on squashed $G_2$ manifolds}

    In this section, we shall study the massless M3-brane whose
transverse space is a deformation of the class of metrics of $G_2$
holonomy whose first-order equations, given in (\ref{simeq}), have
been constructed in \cite{bggg}.  These solutions will provide further
examples of massless M3-brane configurations that are of the same
general pattern as those constructed in \cite{clpmassless}.  We
therefore begin by making the ansatz
\bea
ds_{11}^2 &=& H^2\, dx^\mu\, dx_\mu + d\rho^2 + a^2\, (\td h_1^2 + \td
h_2^2) + c^2\, \td h_3^2 + b^2\, (h_1^2 + h_2^2) + f^2\, h_3^2\,,\nn\\
F_\4 &=& G_\4\,,
\eea
where
\be
h_i\equiv \sigma_i + \Sigma_i\,,\qquad \td h_i = \sigma_i -\Sigma_i\,,
\ee
and $H$, $a$, $b$, $c$ and $f$ are functions of the radial variable
$\rho$ in the 7-metric transverse to the 3-brane world-volume, whose
coordinates are $x^\mu$.  The 4-form $G_\4$ is constructed from
isometry-invariants of the transverse 7-metric.  In terms of the
natural vielbein basis
\be
e^0=h\, dr\,,\quad e^1=a\, \td h_1\,,\quad e^2=a\, \td h_2\,,
\quad e^3 = c\,
\td h_3\,,\quad e^4 = b\, h_1\,,\quad e^5=b\, h_2\,,\quad e^6 = f\,
h_3
\ee
for the metric
\be
ds_7^2=  d\rho^2 + a^2\, (\td h_1^2 + \td
h_2^2) + c^2\, \td h_3^2 + b^2\, (h_1^2 + h_2^2) + f^2\, h_3^2
\ee
in the transverse space, the appropriate invariant ansatz for the
4-form is
\bea
G_\4 &=& u_1\, e^1\wedge e^2\wedge e^4\wedge e^5 + u_2\,
     e^2\wedge e^3\wedge e^5\wedge e^6 + u_2\,
     e^3\wedge e^1 \wedge e^6\wedge e^4 \nn\\
&&+ u_3\, e^0\wedge e^4\wedge e^5\wedge e^6 + u_4\, e^0\wedge
e^1\wedge e^2\wedge e^6 \nn\\
&&+ u_5\, e^0\wedge e^2\wedge e^3\wedge e^4
  + u_5\, e^0\wedge e^3\wedge e^1\wedge e^5\,.
\eea
It is straightforward to calculate $F^2_{ab}$ in the vielbein basis,
leading to
\bea
&&F^2_{00}=6(u_3^2 + u_4^2 +2u_5^2)\,,\quad
F^2_{11}=F^2_{22}=6 (u_1^2 + u_2^2 + u_4^2 +u_5^2)\,,\quad
F^2_{33}=6 (2u_2^2 + 2 u_5^2)\,,\nn\\
&& F^2_{44}=6 (u_1^2 + u_2^2 + u_3^2 + u_5^2)\,,\qquad
F^2_{66}=6 (2u_2^2 + u_3^2 + u_4^2)\,.
\eea

     We parameterise the $u_i$ functions as
\bea
&&u_1=\fft{f_1}{a^2\,b^2}\,,\qquad
u_2=\fft{f_2}{a\,b\,c\,f}\,,\qquad
u_3=\fft{f_3}{b^2\, f}\,,\nn\\
&&u_4=\fft{f_4}{a^2\, f}\,,\qquad
u_5=\fft{f_5}{a\,b\,c}\,.
\eea
The Bianchi identity and equations of motion for the 4-form field
strength in $D=11$ then imply
\bea
dF_4=0:&& f_1'+\ft12 f_3+ \ft12 f_4 -f_5=0\,,\qquad
f_2' +\ft12 f_3 -\ft12 f_4=0\,,\nn\\
d{*F_4}=0:&&(\fft{b^2\,c\,H^4\,f_4}{a^2\,f})' +
\fft{c\, f\, H^4\, f_1}{2a^2\,b^2} -\fft{H^4\, f_2}{c\, f}=0\,,\nn\\
&&\fft{a^2\, c\, H^4\, f_3}{b^2\, f} + \fft{c\, f\, H^4
f_1}{2a^2\, b^2} +\fft{H^4\, f_2}{c\, f}=0\,,\nn\\
&&f_5=\fft{\lambda\, c}{2f\, H^4} -\fft{a^2\, c^2\, f_3}{2b^2\, f^2}
 -\fft{b^2\, c^2\, f_4}{2a^2\, f^2}\,,\label{squashf4eq}
\eea
where $\lambda$ is a constant of integration.  Thus we can solve for
$f_3$, $f_4$ and $f_5$ in terms of the functions $f_1$ and $f_2$,
which satisfy two second-order differential equations.

    A relatively simple way to obtain the equations following from
eleven-dimensional supergravity is to perform a Kaluza-Klein reduction
on the 4-dimensional world-volume of the 3-brane, and hence to
re-express the eleven-dimensional equations in terms of
seven-dimensional ones.  After doing this, it is then a routine
exercise to construct a Lagrangian $L=T-V$ from which, together with
the constraint $T+V=0$, the conditions implied by the
eleven-dimensional supergravity equations may be derived.  We then
look for a superpotential, which will enable us to construct
first-order equations that are first integrals of the original
second-order equations.  In order to find a superpotential for the
system, it seems to be necessary to take the integation constant
$\lambda$ in (\ref{squashf4eq}) to be zero.  A second constant of
integration also needs to be set to zero in order to find a
superpotential, implying that $f_1$ and $f_2$ become algebraically
related:
\be
f_1=\fft{f_2\,(2a\,b\,c- (a^2-b^2)\,f)}{(a^2+b^2)\,f}\,.
\ee
Now our second-order differential equations involve six functions,
namely $a, b, c, f, H$ and $f_2$. We find that their kinetic energy
$T=\ft12 g_{ij}\, (d\a^i/d\eta)\, (d\a^j/d\eta)$ has a  metric whose
inverse is given by $g^{ij}=g_0^{ij} + g_1^{ij}$, where
\be
g_0^{ij} = \fft{1}{18}
\pmatrix{-\ft72 & 1 & 1 & 1 & 1 & 0 \cr
         1 & -\ft72 & 1 & 1 & 1 & 0 \cr
         1 & 1 & -8 & 1 & 1 & 0 \cr
         1 & 1 & 1 & -8 & 1 & 0 \cr
         1 & 1 & 1 & 1& -\ft54 & 0 \cr
         0 & 0 & 0 & 0 & 0 & -\fft{9K_3}{2K_1}}\,,
\ee
and
\be
g_1^{ij} =f_2\, \pmatrix{\ft12 f_2\, Y_2^2 & -\ft12 f_2\, Y_2^2
& -c\, f_2\, Y_2\, Y_3 & c\,f_2\, Y_2\, Y_3 &
0 & -\ft12 f^2\, Y_1\, Y_2\, Y_3
\cr
 -\ft12 f_2\, Y_2^2 & \ft12 f_2\, Y_2^2 & c\, f_2\, Y_2\, Y_3
& -c\, f_2\, Y_2\, Y_3 & 0 & \ft12 f^2\, Y_1\, Y_2\, Y_3
\cr
 -c\, f_2\, Y_2\, Y_3 & c\, f_2\, Y_2\, Y_3 &
2c^2\, f_2\, Y_3^2 & -2c^2\, f_2\, Y_3^2 & 0 &
c\, f^2\, Y_1\, Y_3^2 \cr
c\, f_2\, Y_2\, Y_3 & -c\, f_2\, y_2\, Y_3 &
 -2c^2\, f_2\, Y_3^2 & 2c^2\, f_2\, Y_3^2 & 0 &
 -c\, f^2\, Y_1\, Y_3^2\cr
0 & 0 & 0& 0& 0& 0 \cr
 -\ft12 f^2\, Y_1\, Y_2\, Y_3 & \ft12 f^2\,  Y_1\, Y_2\, Y_3
& c\, f^2\, Y_1\, Y_3^2 &
 -c\, f^2\, Y_1\, Y_3^2 & 0 & -2f^2\, f_2\, Y_1^2}\,,
\ee
where
\crampest
\bea
&&K_1=(3 a^4  + 2 a^2\,  b^2  + 3 b^4 )\, c^2  +
4 a\,b\,c\,f\, (a^2 - b^2) + 4 a^2\,  b^2\,  f\,,\nn\\
&&K_2=(a^2+b^2)^2\, f^2 + 4f_2^2\,,\qquad
K_3=f^2\, (a^2+b^2)\, ((a^4+b^4)\,c^2 + 2a^2\, b^2\, f^2)\,,\nn\\
&&Y_1=c\,(a^4+b^4) + a\, b\, (a^2-b^2)\,,\quad
Y_2=c\,(a^2-b^2) +2a\, b\, f\,,\quad
Y_3=a^2+b^2\,.
\eea
\uncramp

   We find that the potential $V$ can be expressed as $V=-\ft12
g^{ij}\, (\del W/\del \a^i)\, (\del W/\del\a^j)$, where the
superpotential $W$ is given by
\be
W= H^4\,(-a\,  b\, f\, (a^2 + b^2 + c^2) + 1/2 (a^2-b^2)\, c\,
g)\,K\,,
\ee
and
\be
K=\sqrt{1 + 4 f_2^2\,(a^2 + b^2)^{-2}\, g^{-2}}\,.
\ee

        The first-order equations can then be straightforwardly
derived, and are given by
\bea
&&K\,\fft{(a\, H)'}{a\, H}=\fft{a}{4b\, c} -
\fft{b}{4a\, c} -\fft{c}{4a\, b} -\fft{f}{4a^2} +
(K^2-1)\, X\,,\nn\\
&&K\, \fft{(b\, H)'}{b\, H} = -\fft{a}{4b\, c} +
\fft{b}{4a\, c} - \fft{c}{4a\, b} + \fft{f}{4b^2} +
(K^2-1)\, X\,,\nn\\
&&K \fft{(c\, H)'}{c\, H} = -\fft{a}{2b\, c} -\fft{b}{2a\, c}
+\fft{f}{2a\, b} - 2(K^2-1)\, X\,,\nn\\
&&K\fft{(f\, H)'}{f\, H} = \fft{f}{4a^2} - \fft{f}{4b^2} -
2(K^2-1)\, X\,,\nn\\
&&K' = \fft{(K^2-1)(2a\, b\, (a^2+b^2+c^2) - (a^2-b^2)\, c\, f)}{
4a^2\, b^2\, c}\,,\label{eqs222}
\eea
together with $H=K^{1/3}$, and $X$ is given by
\be
X=\fft{a\, b\, (a^2 + b^2 -2 c^2) + (a^2- b^2)\, c\, f}{12a^2\, b^2\,
c}\,.
\ee

  In the previous examples of M3-branes, both in \cite{clpmassless}
and in section 4.1 above, it turns out that by making appropriate
redefinitions of variables, the first-order equations can be reduced
to a subset that are equivalent to the original first-order equations
for the undeformed Ricci-flat metric of $G_2$ holonomy, together with
the additional equations governing the M3-brane metric function $H$
and the functions appearing in the ansatz for the 4-form.  In
particular, this meant that in the M3-brane solution in section 4.1,
the transverse part of the eleven-dimensional metric is of the same
form as the original undeformed Ricci-flat 7-metric of $G_2$ holonomy,
except for ``warp factors'' appearing in the various terms.  In the
situation in the present example, by contrast, it can be seen from
(\ref{eqs222}) that the first-order equations for this M3-brane cannot
be decomposed in such a way that the original first-order $G_2$
holonomy equations (\ref{simeq}) arise as a subset.  Thus in the
present ``squashed'' example, it is not straightforward to obtain an
explicit solution of the M3-brane equations.

\section{Supersymmetry of massless M3-branes}

    One might expect that the existence of superpotentials, and hence
first-order equations for the M3-brane systems, found in the previous
section, would imply that these solutions would be supersymmetric. In
this section, we show that this is indeed the case, for both of the
new classes of M3-branes obtained above.

   The calculations are best performed abstractly, by using the
first-order equations for these solutions.  By this means, one can
establish that in these cases, as in the earlier examples in
\cite{clpmassless}, these first-order equations are precisely the
integrability conditions for the existence of the Killing
spinor.

   In fact an instructive way to study the Killing spinors is
to reverse the logic, and to {\it derive} first-order equations by
requiring the existence of such spinors.  We can then show that these
first-order equations are equivalent to those obtained in sections 4.1
and 4.2 via a construction of the superpotential. In the following two
subsections we shall describe this procedure for the ``round'' and
``squashed'' M3-branes of sections 4.1 and 4.2 respectively.

\subsection{Killing spinor in the ``round'' M3-brane}

   The gravitino transformation rule in $D=11$ supergravity is given by
\be
\delta\hat \psi_A = D_A\, \hat\ep - \fft{1}{288}\, F_{BCDE}\,
\hat \Gamma_A{}^{BCDE}\, \hat\ep +
   \fft{1}{36} \, F_{ABCD}\, \hat\Gamma^{BCD}\,
\hat\ep\,.\label{susytrans}
\ee
It is useful to make a $4+7$ split of the eleven-dimensional Dirac
matrices:
\be
\hat \Gamma_\mu = \gamma_\mu\otimes \oneone\,,\qquad
\hat\Gamma_a= \gamma_5\otimes \Gamma_a\,,
\ee
and to seek Killing spinors of the form $\hat\ep=\ep\otimes\eta$.
Since the field strength $F_\4$ has components only in the
7-dimensional space transverse to the 3-brane, it follows from
(\ref{susytrans}) that in the 3-brane world-volume directions the
requirement $\delta\hat\psi_\mu=0$ leads to $\del_\mu\, \ep=0$,
$\gamma_5\, \ep=\pm \ep$, and the conditions
\be
\ft12 H^{-1}\, \fft{dH}{d\rho}\, \Gamma_0\, \eta \mp \ft{1}{288}\,
F_{abcd}\,\Gamma^{abcd}\, \eta=0\,,\label{mudir}
\ee
respectively,
where, as usual, the ``0'' index denotes the radial direction in the
transverse space.  The matrix that acts on $\eta$ here has zero
eigenvalues if $H$ satisfies one of the following conditions
\bea
\fft{1}{H}\, \fft{dH}{d\rho} &=& \pm \ft16 \sqrt{9u_1^2-9u_4^2
 -(u_2-3u_3)^2}\,,\nn\\
\fft{1}{H}\, \fft{dH}{d\rho} &=& \pm \ft16 \sqrt{u_1^2-u_4^2
 -(u_2+u_3)^2}\,,\label{hpeq1}
\eea
where $u_1$, $u_2$, $u_3$ and $u_4$ are the vielbein components of the
field strength $F_\4$ given in (\ref{f4sf4}), \ie
\bea
&&F_{1245}=F_{2356}=F_{3164} = u_1\,,\qquad F_{0456}= u_2\,,\nn\\
&&F_{0126}=F_{0234}=F_{0315}=u_3\,,\qquad
F_{0156}=F_{0264}=F_{0345}=u_4\,,
\eea
with $u_1=2 f_1/(a^2\, b^2)$, $u_2=f_2/a^3$, $u_3=f_3/(a\, b^2)$ and
$u_4 = f_4/(a^2\, b)$.  (The $\pm$ signs in (\ref{hpeq1}) are not
correlated with those in (\ref{mudir}).)  Specifically, the matrix
acting on $\eta$ has one zero eigenvalue for each sign choice in the
upper equation in (\ref{hpeq1}), and three zero eigenvalues for each
sign choice in the lower equation.  It turns out that it is the upper
equation that is relevant for our purposes.  Either choice of sign in
the upper equation is allowed, since it can be reversed by a change of
orientation conventions.  Equally, an orientation-related sign choice
was possible when we derived the first-order equations in section 4.1.
The convention choices in these first-order equations and in the
Killing-spinor conditions must be appropriately matched if one is to
relate the two.  In fact to obtain agreement with the results in
section 4.1, we should choose
\be
\fft{1}{H}\, \fft{dH}{d\rho} = - \ft16 \sqrt{9u_1^2-9u_4^2
 -(u_2-3u_3)^2}\,.\label{solhp}
\ee

    Since the matrix acting on $\eta$ in (\ref{mudir}) has just one
zero eigenvalue if (\ref{solhp}) is imposed, this means that requiring
$\delta\hat\psi_\mu=0$ determines the Killing spinor $\eta$ in
$D=7$ completely, up to an overall multiplicative function of the
7-dimensional coordinates on the transverse space.  From the
conditions $\delta\hat\psi_a=0$ in the six directions on the $S^3$
bundle over $S^3$ of the principal orbits, we then find that $\eta$
must be independent of the six coordinates on the two 3-spheres.
These components of the gravitino variation also give rise to
first-order equations.  Since the full set of equations are somewhat
involved and cumbersome to present, we shall not give them explicitly
here, but simply record that they eventually turn out to be equivalent
to the first-order equations for the M3-brane solutions that we
obtained in section 4.1.

    Finally, from the component of the gravitino variation in the
radial direction, we can determine the radial dependence of the
Killing spinor.  The result for the case $\gamma_5\, \ep = +\ep$ is
\be
\eta = g_1\, \eta_1 + g_2\,\eta_2\,,\label{g1g2sol}
\ee
where
\bea
g_1&=&H^{1/2}\, (1+\fft{f_3}{b^2})^{1/2}\,
\fft{(a(1-2g)-2{\rm i}\, b)(a+2{\rm i}\, b)}{
\sqrt{(a^2(1-2g)^2 + 4b^2)(a^2 +4b^2)}}\,,\nn\\
g_2&=&H^{1/2}\, (1-\fft{f_3}{b^2})^{1/2}\,
\fft{(a(1+2g)-2{\rm i}\,b)
(a+2{\rm i}\, b)}{\sqrt{(a^2(1+2g)^2 + 4b^2)(a^2 + 4b^2)}}\,,
\eea
and $\eta_1$ and $\eta_2$ are covariantly constant spinors satisfying
the constraint (\ref{roundkscon}).  For $\gamma_5\, \ep=-\ep$, the
associated spinor $\eta$ in the transverse space is again given
by (\ref{g1g2sol}), but now with $f_3$ replaced by $-f_3$.  The
general solution for a Killing spinor can then be written as a
linear combination of $\ep_+\otimes \eta_+$ and $\ep_-\otimes
\eta_-$, where the plus and minus subscripts refer to the two
chirality choices under $\gamma_5$.  Thus in total there will be 4
real solutions, implying $N=1$ supersymmetry in the world-volume of
the M3-brane.  

\subsection{Killing spinor for the ``squashed'' M3-brane}

   We can follow a similar procedure for the ``squashed'' M3-brane
configurations obtained in section 4.2.  Again, one can derive
first-order equations as integrability conditions for the existence of
Killing spinors, and again it turns out that these coincide with the
first-order equations for the M3-brane solutions.  Thus the squashed
M3-branes are also supersymmetric.

  We find that the Killing spinors can again be expressed as a linear
combination of terms $\ep_+\otimes \eta_+$ and $\ep_-\otimes \eta_-$,
where $\gamma_5\,\ep_\pm = \pm\ep_\pm$.  The spinor $\eta_+$ in the 
transverse space is given by
\be
\eta_+ = g_1\, \eta_1 + g_2\,\eta_2\,,
\ee
where
\bea
g_1&=& \sqrt{\fft{H\, (a^2+b^2)\, f}{\sqrt{
4f_2^2 +(a^2+b^2)^2\,f^2)} -2f_2}}\,,\nn\\
g_2&=& \sqrt{\fft{H\, (a^2+b^2)\,f}{\sqrt{
4f_2^2 +(a^2+b^2)\, f^2} +2f_2}}\,,
\eea
and $\eta_1$ and $\eta_2$ are covariantly constant spinors satisfying
the constraint (\ref{roundkscon}).  The spinor $\eta_-$ is given by
the same expressions, but with $f_2$ sent to $-f_2$.  Thus again we
have 4 real solutions in total, and hence $N=1$ supersymmetry in the
world-volume of the M3-brane.

\section{Conclusions}

   The two main results of the paper are (i) the construction of a new
class of metrics with $G_2$ holonomy, and (ii) construction of new
massless M3-brane solutions whose transverse spaces are deformations
of the new $G_2$ metrics.  We then showed that these M3-branes are
supersymmetric. 

    The new class of metrics with $G_2$ holonomy that we found can be
viewed as generalisation of the original $G_2$ metrics of $\R^4\times
S^3$ topology \cite{brysal,gibpagpop}, which have just a single
(trivial) scale parameter, to a new family of {\it two-parameter}
metrics of $G_2$ holonomy, on the manifold of the same $\R^4 \times
S^3$ topology.  This result was obtained by actually starting with a
rather general ansatz for cohomogeneity one metrics whose principal
orbits are $S^3$ bundles over $S^3$.  For the most general ansatz that
we considered, which contains nine functions of the radial coordinate,
we did not find any first-order system of equations as first integrals
of the Einstein equations.  However, for a specialisation to a
3-function ansatz, which is spherically symmetric both in the base and
in the fibre, we were able to find first-order equations derivable
from the superpotential.  The explicit general solution yields metrics
with two parameters (aside from the completely trivial constant
translation of the radial coordinate), one corresponding to the
(trivial) scale size, and the second, called $\lambda$, being
non-trivial and characterising inequivalent metrics.  While for a
generic value of the parameter $\lambda$ these metrics are singular,
they become regular for $\lambda=\{-1,0,+1\}$.  The case $\lambda=0$
corresponds to the original metric of $G_2$ holonomy in
\cite{brysal,gibpagpop}. The solutions possess an $S_3$ automorphism that
allows one to map $\lambda =\{-1,1\}$ to $\lambda =0$, and that allows
the continuum of values for the $\lambda$ parameter to be restricted
to a fundamental domain $0\le\lambda\le\ft13$.

    The second set of results involved explicit solutions for
M3-branes whose transverse spaces are deformations of the
two-parameter metrics of $G_2$ holonomy obtained in setion 2.  We
obtained these M3-brane solutions by first constructing a system of
first-integrals for the equations of $D=11$ supergravity, following
from a superpotential.  We also constructed first-order equations for
M3-branes whose transverse spaces are deformations of the new metrics
of $G_2$ holonomy whose first-order equations were obtained  
in \cite{bggg}.  Just like the M3-branes in
\cite{clpmassless}, these new M3-branes have neither mass nor charge,
and have a naked singularity.  Since this singularity occurs outside
the radius of the innermost ``endpoint'' of the original undeformed
metric of $G_2$ holonomy, this means that the singularities in the
generic Ricci-flat metrics in section 2 do not materially affect the
singularity structure of the associated M3-branes.  We then showed that the
first-order equations in both cases are precisely the integrability
conditions for the existence of a Killing spinor, and hence that the
M3-branes are supersymmetric.

\section*{Acknowledgement}

   We are grateful to Michael Atiyah, Dan Freedman, Sergei Gukov,
Steve Gubser, Nigel Hitchin, Jim Liu, Joe Polchinski and especially Edward
Witten for useful discussions and communications.  M.C. would like to
thank the organizers of the M-theory workshop at the Institute of
Theoretical Physics at the University of California in Santa Barbara
and Caltech theory group for hospitality during the completion of this
work.  C.N.P. is grateful to the high energy theory group at the
University of Pennsylvania and to the Michigan Center for Theoretical
Physics for hospitality during different stages of this work.
Research is supported in part by DOE grant DE-FG02-95ER40893, NSF
grant No. PHY99-07949, Class of 1965 Endowed Term Chair and NATO grant
976951 (M.C.), in full by DOE grant DE-FG02-95ER40899 (H.L.) and in
part by DOE grant DE-FG03-95ER40917 (C.P.).

\section*{Note added}
 
    In an earlier version of this paper, and in \cite{clpmassless}, it
was claimed that the M3-brane solutions were not supersymmetric, but
instead were ``pseudo-supersymmetric'' with respect to a modified
$D=11$ supersymmetry transformation rule.  This incorrect conclusion
resulted from a systematic error in a computer program that we used
for calculating the Killing spinors.  We are grateful to Jim Liu for
calculations that encouraged us to recheck the computer programs and
discover the error.

\end{document}